\documentclass[11pt]{article}
\usepackage{graphicx,amsmath,amsfonts,amssymb}
\usepackage{color,multicol,multirow}
\usepackage{array}
\def\sloppy{\tolerance=100000\hfuzz=\maxdimen\vfuzz=\maxdimen}
\allowdisplaybreaks
\def \beq  {\begin{equation}}
\def \eeq  {\end{equation}}
\def \beqar {\begin{eqnarray}}
\def \eeqar {\end{eqnarray}}
\def\bsp{\beq\begin{split}}
\allowdisplaybreaks
\mathchardef\mhyphen="2D

\def\la {{\langle}}
\def\ra {{\rangle}}

\def\vk {{\vec k}}
\def\vx {{\vec x}}
\def\vy {{\vec y}}

\def\vf {{\varphi}}

\def\Tr {{\rm Tr}}

\def\vk {\vec{k}}
\def\vx {{\vec x}}

\def\vy{\vec{y}}

\def\del {\partial}

\def\A {{\cal A}}

\def\H {{\cal H}}
\def\K{{\cal K}}

\def\O {{\cal O}}

\def\S {{\cal S}}

\def\half{\textstyle{1\over 2}}

\begin{document}
\def \CMP {{Commun. Math. Phys.}}
\def \PRL {{Phys. Rev. Lett.}}
\def \PL {{Phys. Lett.}}
\def \NPBProc {{Nucl. Phys. B (Proc. Suppl.)}}
\def \NP {{Nucl. Phys.}}
\def \RMP {{Rev. Mod. Phys.}}
\def \JGP {{J. Geom. Phys.}}
\def \CQG {{Class. Quant. Grav.}}
\def \MPL {{Mod. Phys. Lett.}}
\def \IJMP {{ Int. J. Mod. Phys.}}
\def \JHEP {{JHEP}}
\def \PR {{Phys. Rev.}}
\begin{titlepage}
\null\vspace{-62pt} \pagestyle{empty}
\begin{center}
\rightline{CCNY-HEP-15/4}
\rightline{July 2015}
\vspace{1truein} {\Large\bfseries
Boundary Conditions as Dynamical Fields}\\
\vskip .1in
{\Large\bfseries ~}\\
\vskip .1in
{\Large\bfseries ~}\\
{\large\sc Dimitra Karabali$^a$} and
 {\large\sc V.P. Nair$^b$}\\
\vskip .2in
{\itshape $^a$Department of Physics and Astronomy\\
Lehman College of the CUNY\\
Bronx, NY 10468}\\
\vskip .1in
{\itshape $^b$Physics Department\\
City College of the CUNY\\
New York, NY 10031}\\
\vskip .1in
\begin{tabular}{r l}
E-mail:&{\fontfamily{cmtt}\fontsize{11pt}{15pt}\selectfont dimitra.karabali@lehman.cuny.edu}\\
&{\fontfamily{cmtt}\fontsize{11pt}{15pt}\selectfont vpn@sci.ccny.cuny.edu}
\end{tabular}

\fontfamily{cmr}\fontsize{11pt}{15pt}\selectfont
\vspace{.8in}
\centerline{\large\bf Abstract}
\end{center}
The possibility of treating boundary conditions in terms of a bilocal dynamical field
is formalized in terms of a boundary action. This allows for a simple path-integral perturbation theory approach to physical effects such as radiation from a time-dependent boundary.
The nature of the action which governs the dynamics of the bilocal field is investigated for
a limited case (which includes the Robin boundary conditions).

\end{titlepage}

\pagestyle{plain} \setcounter{page}{2}
\setcounter{footnote}{0}
\setcounter{figure}{0}
\renewcommand\thefootnote{\mbox{\arabic{footnote}}}
\fontfamily{cmr}\fontsize{11pt}{15pt}\selectfont

\section{Introduction}
The role of boundary conditions in field theory has recently been of considerable research interest.
This has been explored in great detail in the case of the Casimir effect, for a variety of 
geometries \cite{reviews}.
The question 
obviously goes beyond the Casimir effect, which only pertains to the vacuum-to-vacuum amplitude.
The impact of different boundary conditions on higher point functions is clearly of importance in diffraction theory and other physical phenomena.
The general theory of self-adjoint extensions \cite{{vonNeumann}, {asorey}} also allows for boundary conditions with negative eigenvalues for the operators of interest such as the Laplace operator; these are related to edge excitations and can lead to
interesting phenomena \cite{{bal}, {govindarajan}}.

From the mathematical point of view, boundary conditions are necessary to render the problem
well-defined with the needed self-adjointness properties. But from a physical point of view,
boundary conditions are idealizations of dynamics on the boundary. Thus the dynamics of material particles which constitute a metal plate would lead to nonzero electrical
conductivity and, in an idealized limit, would give
the standard Dirichlet and Neumann conditions used in Casimir calculations
with the electromagnetic field. Thus it is physically meaningful to have boundary conditions which can vary from point to point on the boundary or which can evolve with time, in accordance with
some dynamical principle.
Effectively, one must treat boundary conditions as additional dynamical fields with their own action and time-evolution.
In particular, one may ask how the dynamics of boundary conditions for 
a given set of fields is modified by the back reaction of the
fields themselves.
In this paper, we start exploring some of these questions.

In considering diffractive contributions  to the Casimir effect, we have recently developed 
a boundary action approach \cite{KKN1}-\cite{KN1}.
The strategy is to obtain a lower dimensional
field theory defined on the boundaries, as a functional of the boundary value of the field,
by integrating over the bulk fields. 
This boundary action makes it straightforward to incorporate
the effects due to edges and apertures on the boundary
as part of the integration over the boundary values of the fields.
The formalism was also extended to include general boundary conditions as allowed by the von Neumann theory of self-adjoint extensions \cite{KN1}.
In this case, an integral kernel in the boundary theory encodes the allowed
boundary conditions.
This formulation
has the advantage of recasting the entire discussion of boundary conditions and how they can change with space or time as a lower dimensional field theory which lives on the boundary.
In particular, one does not have to modify the mode expansions of the bulk fields as 
one changes the boundary conditions.
Thus boundary conditions can indeed be treated as 
a dynamical field.
In this paper, we will take the next logical step 
 and consider how the bulk fields can affect the dynamics of the boundary conditions.
 
 The paper is organized as follows. In the next section, we will briefly review the boundary action. 
 In section 3, we consider radiation from the boundary due to
 time-dependent Robin conditions. This problem is closely related to the dynamical Casimir effect since time-dependent boundary conditions can simulate moving mirror effects \cite{experiment} - \cite{fosco}.
The calculation of the radiation from boundaries with time-dependent Robin boundary conditions was done by using Bogoliubov transformations to define
a new set of creation and annihilation operators for the field \cite{{SF}, {Rego}}.
Our method will be much simpler, involving the perturbative expansion of a path-integral for boundary fields, which can easily accommodate arbitrary dimensions, higher order effects, corrections due to bulk interactions of fields, etc.
 
 In section 4, we consider the effective action for boundary conditions, i.e.,
 for the integral kernel or the bilocal field on the boundary which encodes the boundary conditions, which obtained
 by integrating out
 the fields. This is a rather involved problem, our calculation is for 
 a limited choice of the bilocal field.
Nevertheless, it illustrates the method and some general features of the action
for this field. A more complete analysis is clearly called for, this is currently under investigation.

The paper concludes with a short discussion.

\section{ General boundary conditions}

We review the boundary action briefly considering a free scalar field
$\phi$ in a region $V$ with the usual kinetic term of the form
$\int_V (\del \phi )^2/2$. On the boundary, the combination
$\vf + i \del_n \vf$, where $\vf$ is the boundary value of
$\phi$ and $\del_n \vf$ is its normal derivative, is to be viewed as an
element of a Hilbert space of ${L}^2$-functions.
The general boundary condition, according to the von Neumann theory
of self-adjoint extensions, is \cite{vonNeumann, asorey}
\beqar
\vf +i \,\del_n \vf &=& U \, ( \vf - i \,\del_n \vf )\nonumber\\
(\vf +i \,\del_n \vf) (x) &=& \oint_y U(x,y)  \, ( \vf - i \,\del_n \vf )(y)\label{1}
\eeqar
where $U$ is a unitary operator on the boundary Hilbert space; this is made more explicit
in the second line by writing $U(x, y)$.
We can also rewrite the boundary condition as
\beq
\del_n \vf = -i \left( {U -1 \over U+1}\right)\, \vf \equiv - {\cal K} \, \vf \label{2}
\eeq
where ${\cal K}$ is a hermitian operator; it corresponds to the Cayley transform of
$U$.
The limit ${\cal K} \rightarrow 0$, equivalent to $U =1$, gives the Neumann
boundary condition, while ${\cal K} \rightarrow \infty$ gives the Dirichlet condition,
as seen by dividing (\ref{2}) by ${\cal K}$ and taking the limit ${\cal K} \rightarrow \infty$.
The case of ${\cal K}$ being a constant (proportional to the identity
on the Hilbert space) is the Robin condition.
These are special points on the space of boundary conditions;
clearly, in general,
 there is a much larger class of choices.

In \cite{KKN1}, we considered 
a two-step evaluation of the Euclidean partition function.
While variants of this, including the possibility of interacting fields were also considered
in \cite{KKN2}, for the present discussion, we will consider a free massless scalar field theory
with a standard action $S(\phi ) = \half  \int (\del \phi )^2$.
Further we consider a boundary which is a plane normal to the $x_1$ direction.
We can then write the partition function as
\beq
Z = \int [d\phi] \, e^{- S (\phi )}
\label{3}
\eeq
with the field $\phi (x)$ parametrized as
\beq
\phi (x) = \int_{x'} \, \vf (x') \, n \cdot \del_{x'} G(x', x) ~+~ \eta (x)
\label{4}
\eeq
where $G(x', x)$ is the Green's function for the Laplacian in the bulk with Dirichlet boundary conditions,
$\vec n$ is the unit vector normal to the boundary
and $\vf$ is the boundary value of $\phi$. The field $\eta (x)$ also obeys the Dirichlet condition
$\eta = 0 $ on the boundary. The parametrization
(\ref{4}) ensures the appropriate boundary behavior for $\phi$.
Integrating out $\eta$ one obtains
\beq
Z = \det (- \square )_V ~ \int [d\vf ] \, e^{- S_{B, \vf}  (\vf )}
\label{5}
\eeq
where $S_{B, \vf}$ denotes the boundary action. It is of the form
\beqar
S_{B, \vf} &=& {1\over 2} \int_{x', y'} \vf(x') \, M(x', y' )\, \vf (y')\nonumber\\
M(x', y') &=& n\cdot \del_{x'} \, n\cdot \del_{y'} G(x', y')
\label{6}
\eeqar
The general boundary condition (\ref{2}) can be incorporated by using the augmented boundary action
\cite{KN1}
\beq
\S_{B} (\vf, \K ) = {1\over 2} \int_{x', y'} \vf(x') \, \left[M(x', y' ) + \K (x', y') \right]\, \vf (y') 
\label{7}
\eeq
The parametrization (\ref{4}) does not imply any boundary condition on $\phi$ since the boundary value $\vf$ is unrestricted. One may even think of the plane as a fictitious plane where we have $\phi = \vf$. The boundary conditions on $\phi$ are imposed by the boundary action (\ref{7}), when we integrate over $\vf$, after the choice of $\K$. As said before, $\K=0$ corresponds to Neumann and $\K \rightarrow \infty$ to Dirichlet conditions.

To see how action (\ref{7}) comes about, we start by considering the normal derivative of the
field. 
Consider formulating the functional integral for $Z$ by discretizing the coordinate along the 
normal direction, say, the $x_1$-direction. The Euclidean action which enters
(\ref{3}) then has the form
\beq
S = S( \{ \phi_i \}) = {1\over 2} \int d^3x^T~ \left[ {(\phi_N - \phi_{N-1})^2\over x_N - x_{N-1}}
+ {(\phi_{N-1} - \phi_{N-2})^2\over x_{N-1} - x_{N-2}} + \cdots + (\nabla^T\phi)^2\right]
\label{7a}
\eeq
where
$\phi_N = \vf $ is the boundary value of $\phi$ and the superscript $T$ indicates components tangential to the boundary. Integrating out the fields
$\phi_i$, keeping $\phi_N = \vf$ fixed, we get the boundary action
$S_B (\vf)$ given as
\beq
e^{-S_B(\vf)} = \int \prod_1^{N-1} d\phi_i  \, ~\exp (- S (\vf, \{\phi_i \})
\label{7b}
\eeq
If we functionally differentiate $e^{-S}$ with respect to
$\vf = \phi_N$, we find
\beq
{\delta \over \delta \vf} ~e^{-S} \equiv {\delta \over \delta \phi_N} ~e^{-S} = -{(\phi_N - \phi_{N-1} ) \over x_N - x_{N-1}}~e^{-S}\,
\rightarrow - \del_n \vf ~ e^{-S}
\label{7c}
\eeq
In other words, the normal derivative can be obtained as the result of functional differentiation
of the $e^{-S}$ with respect to the boundary value of the field.
This is a key result for us as it gives a way to express the normal derivative entirely in terms of the boundary action.

We can now see that the general boundary condition (\ref{2}) can be obtained for the remaining integration over $\vf$ if we use the augmented action $\S_B (\vf, \K)$ from
(\ref{7}). For this consider the identity
\beqar
 \int [d\vf] \, \exp\left( - \S_B(\vf, \K) \right) \, (\del_n \vf)&=&
 \int [d\vf] \, \exp\left( {-{1\over 2} \int \vf\, {\cal K}\,  \vf } - S_B (\vf) \right) \, (\del_n \vf)\nonumber\\
&=&  \int [d\vf] \, \exp\left( {-{1\over 2} \int \vf\, {\cal K}\,  \vf }\right) \,\, \left( -{\delta \over \delta \vf}\right)\,e^{-S_B(\vf)}\nonumber\\
&=& \int [d\vf] \, \exp\left( {-{1\over 2} \int \vf\, {\cal K}\,  \vf }\right) \,(- {\mathcal K}\, \vf )\, e^{-S_B(\vf)}
\label{7d}
\eeqar
where, in the last line, we have done a partial functional integration. 
The first and last steps of this equation show that
\beq
 \int [d\vf] \, e^{- \S_B (\vf, \K) } \,\, (\del_n \vf  + {\mathcal K}\vf)
= 0 \label{7e}
 \eeq
This justifies our argument that the augmented boundary action (\ref{7}) incorporates the general boundary condition (\ref{2}).

It is worth pointing out that the use of the action (\ref{7}) to take care of boundary conditions has 
some advantageous features. 
The field $\eta$ and the Green's function $G(x', x)$ in (\ref{4}) obey a fixed boundary condition, namely, the Dirichlet condition. The real genesis of various boundary conditions for the field $\phi$
 is transferred to
$\vf$ and the action (\ref{7}) which controls it. This makes it much easier to analyze change of boundary conditions, including time-dependence, dynamical determination of boundary conditions, how 
back-reaction from other fields can modify the boundary conditions, etc.
We can treat the boundary conditions effectively as a boundary field. The fact that this
approach is viable, namely that the boundary action (\ref{7}) does reproduce the physical effects of different boundary conditions correctly, is clear from calculations of the Casimir energy \cite{KN1, cavalcanti}. We also point out that a similar idea of dynamically implementing the standard Robin boundary condition in terms of a boundary action was also suggested in \cite{cavalcanti}, although our formalism is somewhat more general.

Going back to (\ref{7}), we see that we can impose different conditions on different subsets of
the boundary. Regions with Neumann condition will have $\K =0 $, while
regions on the boundary where ${\cal K} \rightarrow \infty$ must have
$\vf =0$.
We may regard the hermitian operator ${\cal K}$ as an integral kernel, as already
indicated
in (\ref{7}). Thus, generally, boundary conditions correspond to a {\it bilocal field}
${\cal K}(x', y')$ on the boundary and the general form of the boundary action is
\beq
\S_B = \S_{B}(\vf, \K) + S_{\cal K}\label{8}
\eeq
where $S_{\cal K}$ is the action for the field ${\cal K} (x', y' )$.
We have obtained the form of $\S_B (\vf, \K)$, but the question arises: What is the form
of the action $S_\K$ for the bilocal field $\K (x, y)$? A related question is: To what extent is it advantageous to think of $\K (x, y)$ as a dynamical field?

We will take up the second question first.
In the next section, we consider some examples
which highlight the
value of viewing $\K$ as a field. We calculate the
radiation from a time-dependent boundary condition.
Such boundary conditions can be engineered in many ways: For example, 
for the electromagnetic field, one can consider a metal plate as
the boundary and modulate the conductivity. This can also be viewed as
the problem of moving mirrors where the effect of the movement can be transferred to 
time-dependence of the boundary condition \cite{experiment}.

Regarding the first question, clearly, 
the form of $S_\K$ is to be determined ultimately by the physics of the material of the boundary.
However, we can gain some insight
into the nature of $S_\K$ by calculating corrections to it due to the fields
$\vf$, i.e., by integrating out $\vf$. This is familiar from standard field theory. If we integrate out one set of fields and obtain the corrections to the action for another set of fields, we can postulate that the action for the latter must at least have terms of the appropriate structure
(as monomials of the fields and their derivatives) to provide proper renormalization for the
terms generated by integrating out the first set of fields.
A simple and age-old example is when we consider fermions coupled to a scalar field. Integrating out 
the fermions shows that we must have the usual kinetic term, mass term and an
additional quartic self-coupling for the scalar
for proper renormalization.
This is the strategy we will follow here. We will consider the nature of the action for
$\K$ generated
by integrating out the $\vf$ fields.

Finally, although we will use a simple free field theory example for the calculations to follow,
we may note that the formulation is quite general and can easily handle interacting theories.
The construction of the boundary action for the general case was briefly discussed in
\cite{KKN2}. The part of the boundary action without the $\K \vf^2$ term
is defined by
\beq
e^{-S_B (\vf )} = \int [d\eta] \, e^{- S(\eta + \vf )}
\label{8a}
\eeq
where $\eta$ vanishes on the boundary and $\phi$
takes the value $\vf$ on the boundary.
$\phi$ is extended into the bulk from its value $\vf$ on the boundary in a unique way, so that
there are no additional functional degrees of freedom associated with it.
The result (\ref{8a}) can be phrased as follows.
Consider the generator $\Gamma [\chi ]$
of the 1PI vertices of the theory, which is defined by
\beq
\exp \left( - \Gamma [\chi ]  \right)
= \int [d\eta ] ~\exp \left( - S[\chi + \eta] + \int {\delta \Gamma \over \delta \chi} \eta
\right)
\label{8b}
\eeq
Here we will consider evaluating $\Gamma$ such that the field $\eta$ vanishes
on the boundary. This means that the Green's functions occurring  in, say,
 a perturbative expansion of
$\Gamma$ will obey the Dirichlet conditions.
So far the field $\chi$ is an arbitrary field. We now choose $\chi$ 
as a particular solution of the equation of motion
\beq
{\delta \Gamma \over \delta \chi} = 0
\label{8c}
\eeq
obeying the condition
$\chi \rightarrow \vf $ on the boundary. Comparing (\ref{8a}) and (\ref{8b}) with this condition,
we see that the integral in (\ref{8b}) gives the boundary action.
In other words,
\beq
S_B [\vf ] = \Gamma [ \chi ],  \hskip .2in {\rm subject~ to}~~{\delta \Gamma \over \delta \chi } =0,
\hskip .2in {\rm and}~~\chi \rightarrow \vf ~~{\rm on~ the~ boundary}
\label{8c}
\eeq
The full boundary action can then be written as
\beq
\S_B =  \Gamma [\chi ] \Biggr]_{(\delta \Gamma/\delta \chi)= 0, \, \chi\rightarrow
\vf} ~+~ {1\over 2} \int_{x', y'} \vf (x') \, \K (x', y' ) \, \vf (y') ~+~ S_\K
\label{8d}
\eeq
This gives the boundary action including the effect of interactions.

\section{Radiation from boundary}

We will consider a single flat boundary, say at $x_1 =0$, with
Robin boundary conditions for the field $\phi$,
with the parameter $\kappa$ taken to be a local boundary field.
In other words, we take $\K (x , y) = \kappa(x)\, \delta^{(3)} (x - y)$. We use a notation where the coordinates of a point in the bulk are denoted by $(x, x_1)$, where $x$ corresponds to directions tangential to the boundary surface, including the Euclidean time direction and $x_1>0$ is along the direction  perpendicular to the boundary.
One immediate consequence of time-dependent boundary conditions is radiation from the boundary; this has been calculated
for scalar fields and time-dependent Robin boundary conditions in (1+1) dimensions \cite{SF} by carrying out a Bogoliubov transformation on the fields.
From our point of view, the Robin boundary condition corresponds to a boundary action term
$\int \kappa\, \vf^2 /2$ for a scalar field. Thus we may view the radiation as a decay process
for the $\kappa$ field, $\kappa \rightarrow \vf \, \vf$. It is then straightforward to calculate this in simple perturbation theory.
The physical observation is that we have two detectors placed somewhere in the bulk,
say at points $x$ and $y$,
which show the transition corresponding to the absorption of two of the $\vf$-particles.
The boundary fields propagate into the bulk as given in
(\ref{4}). Let $J$ designate a source operator associated with
the detector whose matrix element effects the transition detecting the $\vf$-particle.
The amplitude for the process is then given by
\beqar
{\mathcal A} &=& \int J_{fi}(x,x_1)\, J_{f'i'}(y, y_1) ~\la \phi(x,x_1) \phi(y,y_1) \ra~ \nonumber\\
&=& \int J_{fi}(x,x_1) J_{f'i'}(y, y_1) \la \vf(x') \vf(y') \ra
\del_{1'} G(x',x'_1; x,x_1)|_{x'_1=0} \, \del_{1'} G(y', y'_1; y,y_1)|_{y'_1=0} 
\label{9}
\eeqar
where $\la \vf(x') \vf(y') \ra$ is the propagator for $\vf$ as given by the boundary action
and $G(x,x_1; y,x_1)$ is the Dirichlet Green's function. In order to derive the transition amplitude (\ref{9}) corresponding to the absorption of two particles of energy $\omega$ and $\omega'$, it is necessary to analytically continue to Minkowski space. We briefly outline how this is done. For a single boundary, taken to be at $x_1 = 0$, the Euclidean Green's function $G_E$ is given by
\beq
G_E(x,x_1;y,y_1)= 2 \int {d^d k \over (2 \pi)^d} \int_0^{\infty} {dk_1 \over \pi}{{e^{i k (x-y) }} \over {k^2 +k_1^2} } \sin(k_1 x_1) \sin(k_1 y_1)
\label{12a}
\eeq
The analytical continuation to Minkowski signature is done by using the substitutions
\beqar
x_E &= & (x_{d+1}, \vec{x}_{||}) \rightarrow x_M=(ix_0, \vec{x}_{||}) \nonumber \\
k_E & = & (k_{d+1}, \vec{k}_{||}) \rightarrow k_M=(-ik_0, \vec{k}_{||})
\label{12b}
\eeqar
and the $i \epsilon$ prescription. Here $x_{d+1}$ is the Euclidean time and $\vec{x}_{||}$ denotes the spatial directions tangential to the boundary. Using (\ref{12b}) we find
\beq
\del_{x_1} G(x,x_1;y,y_1)|_{x_1=0}  =  -i \int d\mu( k) \int _{-\infty}^{\infty}  {dk_1 \over {i\pi}}  {{e^{i k (x-y)} k_1e^{ik_1 y_1} }\over {k_1^2+\vec{k}_{||}^2-k_0^2-i\epsilon} } 
\label{12c}
\eeq
where $d\mu (k) = {{dk_0 d\vec{k}_{||}} / {(2 \pi)^d}}$. Using contour integration we find
\beq
\del_{x_1} G(x,x_1;y,y_1)|_{x_1=0}  =  -i \int d\mu( k)   e^{i k (x-y)} e^{ik_1 y_1} 
\label{12d}
\eeq
where 
\beq
k_1= \left \{ 
\begin{matrix}
\sqrt{k_0^2 -\vec{k}_{||}^2} & ~~{\rm if} ~~~~k_0^2 > \vec{k}_{||}^2 \\
i \sqrt{ \vec{k}_{||}^2-k_0^2} & ~~{\rm if} ~~~~ \vec{k}_{||}^2 > k_0^2  \\
\end{matrix}
\right.
\label{dG}
\eeq
Further
\beq
M(x,y) = \del_{x_1} \del_{y_1} G(x,x_1;y,y_1)|_{x_1,y_1=0} = -i \int d\mu (k) e^{i k (x-y)} i k_1
\label{M}
\eeq
where $k_1$ is given in (\ref{dG}). 
The Minkowski boundary action for the field $\vf$ is thus given by
\beq
S_B(\vf, \kappa) = -{1\over 2} \int \vf (x) \left[ ik_1 + \kappa \right]_{x,y} \vf(y)= -{1\over 2} \int \vf (x) \left[ ik_1 + \kappa_0 + \delta \kappa (x) \right]_{x,y} \vf(y)
\label{10}
\eeq
where we have separated out a constant term $\kappa_0$
which is space-time independent. The last term in (\ref{10}) can be treated as the interaction part in a perturbation scheme. The propagator for $\vf$ is then given by
\beqar
\la \vf(x) \vf(y) \ra &=& \la \vf(x) \vf(y) \ra_{\kappa_0} +  \int_{z} 
 \la \vf(x) \vf(z) \ra_{\kappa_0} (-i \delta \kappa(z)) \la \vf(z) \vf(y) \ra_{\kappa_0} \nonumber\\
 &+ & \int_{z} \int_{z'}
 \la \vf(x) \vf(z) \ra_{\kappa_0} (-i \delta \kappa(z) ) \la \vf(z) \vf(z') \ra_{\kappa_0} (-i \delta \kappa(z')) \la \vf(z') \vf(y) \ra_{\kappa_0} 
 + \cdots
\label{11}
\eeqar
where 
\beq
 \la \vf(x) \vf(y) \ra_{\kappa_0}= -i \int d\mu (k)  {{e^{i k (x-y) }} \over {\kappa_0+ik_1} }
\label{12}
\eeq

Using (\ref{12d}) and (\ref{11}) we can in principle calculate the transition amplitude in (\ref{9}) in arbitrary dimensions and to all orders in $\delta \kappa$. The calculation is particularly straightforward in (1+1) dimensions. In this case we find that the transition amplitude for the absorption of particles of energy $\omega,~\omega '$ by the detectors is given by 
\beqar
{\mathcal A} & = & i ~
{{J_{fi}(\omega, |\omega| )  J_{f'i'}(\omega', |\omega'|)}   \over {(\kappa_0 +i |\omega | )
(\kappa_0 +i |\omega'|)} } ~\Bigl[ - \delta \kappa (-\omega - \omega')
+\int {d\xi^{(1)} \over {2\pi}} {{\delta \kappa (-\omega - \xi^{(1)})\delta \kappa (\xi^{(1)} - \omega')} \over (\kappa_0 +i |\xi^{(1)} | )} \nonumber \\
&&\cdots -(-1)^n \int {d\xi^{(1)} \over {2\pi}} \cdots {d\xi^{(n)} \over {2\pi}}{{\delta \kappa (-\omega - \xi^{(1)})\delta \kappa (\xi^{(1)} - \xi^{(2)})\cdots \delta \kappa (\xi^{(n)} - \omega')} \over {(\kappa_0 +i |\xi^{(1)}| )\cdots (\kappa_0 +i |\xi^{(n)} | )}}  \Bigr]
\label{13}
\eeqar
where $J$ and $\delta \kappa$ in (\ref{13}) refer to the appropriate Fourier transforms.
In order to calculate the frequency spectrum of the particle number observed in the detectors
we have to square the amplitude and sum over the final states. The last step is a bit tricky
since we are summing over the final states of the detector. For a perfect detector,
this process must be equivalent to integrating over the final particle phase space.
This means that for the summation over a small range of final states we can use
\beq
\sum_f  J_{fi} (\omega ) J^{*}_{fi} (\omega ) = 2 \omega {d\omega \over 2 \pi} 
\label{14}
\eeq
A more detailed argument justifying (\ref{14}) is given at the end of this section. Using this and integrating over $\omega'$ we obtain the particle number distribution as observed in one detector as
\beqar
dN(\omega) &=&  d \omega\, {2 \omega \over \pi}
\int {d \omega'\over 2 \pi} \, { {\omega' \,\Theta (\omega')} \over {(\kappa_0^2 + \omega^2 ) ( \kappa_0^2 + \omega'^2)}}\times\nonumber\\
&& {\Bigg\vert} \biggl[-\delta \kappa (-\omega - \omega') +\int {d\xi^{(1)} \over {2\pi}} {{\delta \kappa (-\omega - \xi^{(1)})\delta \kappa (\xi^{(1)} - \omega')} \over (\kappa_0 +i |\xi^{(1)} | )} \nonumber \\
&&\cdots -(-1)^n\!\! \int {d\xi^{(1)} \over {2\pi}} \cdots {d\xi^{(n)} \over {2\pi}}{{\delta \kappa (-\omega - \xi^{(1)})\delta \kappa (\xi^{(1)} - \xi^{(2)})\cdots \delta \kappa (\xi^{(n)} - \omega')} \over {(\kappa_0 +i |\xi^{(1)}| )\cdots (\kappa_0 +i |\xi^{(n)} | )}}  \biggr]{\Bigg\vert}^2~~
\label{15}\\
&=& d \omega\, {2 \omega \over \pi}
\int {d \omega'\over 2 \pi} \, { {\omega' \,\Theta (\omega')} \over {(\kappa_0^2 + \omega^2 ) ( \kappa_0^2 + \omega'^2)}}~{\big\vert} T(-\omega -\omega'){\big\vert}^2
\label{15T}
\eeqar
Expression (\ref{15}) is in agreement with previous results for particle production rate from time-dependent Robin boundary conditions. In \cite{SF} the frequency resolved particle production rate was calculated using Bogoliubov transformation for the fields up to second order in $\delta\kappa$. This agrees with (\ref{15}) with the notational identification,
$\kappa = -1/\gamma$, $\kappa_0 = -1/\gamma_0$, $\delta \kappa = -\delta \gamma /\gamma_0^2$. Some higher order corrections, argued to be relevant in experimental setups for detecting dynamical Casimir effect \cite{experiment},  were calculated in \cite{Rego}, again using Bogoliubov transformations. Our expression (\ref{15}) is in agreement with these results as well, after the appropriate translation in the notation. Finally, we note that the term in the square brackets in
(\ref{15}) is the $T$-matrix if one regards $\delta \kappa$ as a potential energy term for the 
$\vf$ on the boundary. This is made explicit in (\ref{15T}).

The approach we used in calculating the radiation intensity is quite general and is applicable in arbitrary dimensions. We will now outline the corresponding derivation in $(3+1)$ dimensions to order $(\delta \kappa)^2$. In this case $\delta \kappa (x,t)$ is in general a function of both space and time. Following the same procedure as before, we find that the transition amplitude for an absorption of two particles of frequency $\omega$ and $\omega'$ by the detectors when they are placed far away from the boundary is
\beq
{\mathcal A}  =  -i ~
{J_{fi}(\omega, \vec{k}_{||}, k_1 )  J_{f'i'}(\omega', \vec{k'}_{||}, k'_1)} \,\, {\delta \kappa (-\omega - \omega', -(\vec{k}+\vec{k'})_{||}) \over {(\kappa_0 +i |k_1 | )
(\kappa_0 +i |k'_1|)} } 
+ \cdots
\label{16}
\eeq
where $k_1= \sqrt{\omega^2-\vec{k}_{||}^2} $ and $k'_1= \sqrt{\omega^{'2}-\vec{k'}_{||}^{2} }$. 
This is the far-field contribution. If the detectors are placed very near the
boundary, there will be a near-field contribution. The latter corresponds to the
contribution from the range of $\vec{k}_{||}$ such that $\vec{k}_{||}^2 > \omega^2$.
We are ignoring this for detectors placed far from the boundary. (The distinction between what is far and what is near is controlled by the wave length of the particle.)
To continue along the lines of (\ref{16}), the summation over the final states gives

\beq
\sum_f \, J(\omega, \vk ) J^*(\omega, \vk ) = 2\, \omega_k \, {d^3 \vec{k} \over (2 \pi )^3}
\label{14d}
\eeq
This is the higher dimensional analogue of (\ref{14}).
Squaring the amplitude (\ref{16}), using (\ref{14d}) and integrating over $\omega'$ with $k'_1= \sqrt{\omega^{'2}-\vec{k'}_{||}^{2} }$ we find
\beq
dN(\omega) = {{2 \omega}} { d^3 \vec{k}\over (2\pi)^3} \int _{|\vec{k'}_{||}|} ^{\infty} {{\omega^{'2} d \omega'} \over { \pi \sqrt{\omega^{'2}-\vec{k'}^{2}_{||}}}}~ {{\big\vert} \delta \kappa (-\omega - \omega', -(\vk+\vec{k'})_{||}) {\big\vert}^2\over {(\kappa_0^2 +\omega^2-\vec{k}^2_{||} )\, (\kappa_0^2 +\omega^{'2}-\vec{k'}^{2}_{||} )}}~ {d^2\vec{k'}_{||} \over (2\pi)^2}
\label{15a}
\eeq
If $\delta\kappa (t)$ depends only on time and is space independent, (\ref{15a}) 
can be simplified further. We find
\beq
dN(\omega) = A~ {{2 \omega} } {d^3 \vec{k} \over {(2\pi)^3}} \int _{|\vec{k}_{||}|} ^{\infty} {{ \omega^{'2} d \omega'} \over { \pi \sqrt{\omega^{'2}-\vec{k}^{2}_{||}}}} {\vert \delta \kappa (-\omega - \omega') \vert^2\over {(\kappa_0^2 +\omega^2-\vec{k}^2_{||} )(\kappa_0^2 +\omega^{'2}-\vec{k}^{2}_{||} )}} 
\label{15b}
\eeq
where $A$ is the area of the boundary.

The basic idea in demonstrating the calculations
(\ref{15}) and (\ref{15a}, \ref{15b}) was to phrase everything as a standard perturbative
calculation in field theory, so that it brings out the nature of $\kappa$ as a boundary field;
it also shows how one can easily include corrections, due to back-reaction from the bulk fields
as well as multiple emission processes.

Before we close this section, for completeness,
we give the argument justifying
(\ref{14}) and (\ref{14d}).
These are really a standard result although not expressed in this form in most calculations.
The amplitude for the emission (or absorption or scattering) of particles is usually written in the form of 
the integral over all coordinates of the product of a vertex function $V(x_1, x_2\cdots)$ 
and the single particle wave functions, 
\beq
\A = \int_{\{ x\} } V(x_1, x_2, \cdots, x_n, x) \, u_{k_1} (x_1) \, u_{k_2}(x_2)
\cdots {e^{i \vec{k} \cdot \vec{x}} \over \sqrt{2 \omega_k V} }
= \int_x F(x) \, {e^{i \vec{k} \cdot \vec{x}} \over \sqrt{2 \omega_k V} }
\label{14a}
\eeq
where $u_k$'s are the wave functions (or their conjugates as needed),
and in the second expression we have abbreviated the integral since we want to focus on one emitted particle of momentum $\vk$.
The field is taken to be free and enclosed in a spatial cubical box of volume $V$ with, say, periodic boundary conditions (with $ V \rightarrow
\infty$ eventually).
The summation over final states in the square of this amplitude gives the factor
$V (d^3 k/(2\pi)^3)$, so that the phase space measure from the particle under consideration
is
\beq
{1 \over 2 \omega_k V } \, V {d^3 \vk \over (2 \pi)^3} = {1 \over 2 \omega_k  } \,  {d^3 \vec{k} \over (2 \pi)^3}
\label{14c}
\eeq

There is another way to think about this process. The particles are emitted from the interaction region
and absorbed by some detectors. One can consider this whole process; in fact, this is even more physical as this is exactly what is done in any experimental observation.
The emitted particle propagates from the boundary to the detector (taken to be at spacetime point $y$, with $y^0 > x^0$)
and so the amplitude for this is
\beqar
\A &=& \int_{x,y} F(x)\, G(x, y) \, J(y) = \int_{x,y} F(x)\, {e^{ip_0(x_0-y_0)-i\vec{p}\cdot (\vec{x}-\vec{y}))}\over 2 \omega_p }
\, {J(\omega, \vk ) \over 2 \pi} \, e^{i \omega y^0 - i \vk \cdot \vy}\nonumber\\
&=& \int_x F(x)\, e^{i \vk \cdot \vx} \,\, {J(\omega, \vk ) \over 2 \omega_k}
\label{14b}
\eeqar
Upon squaring the amplitude (\ref{14b}) and summing over a small range of final states
(which are now the final states of the detector), we get
a factor $\sum_f \vert J \vert^2 / 4  \omega_k^2$. 
This factor will be sensitive to the detection efficiency via the $\vert J \vert^2$, but
we can consider a perfect detector where this should give the same result as we obtained for the free particle.
The agreement of this factor
with (\ref{14c}) then shows that we need (\ref{14d})
for a perfect detector. 

This second way of thinking about the process has some advantage for us.
For the first method, we will need wave functions with correct boundary conditions with $\K$ included.
While this is not particularly difficult, our approach has been to use the boundary action directly,
so it is interesting to bypass the wave functions. 
The formula (\ref{14d}) re-expresses the needed summation as a property of the detector, and so the same result can be used for our case as well. The same detector can be used
for the particles emitted by the boundary.
(Strictly speaking there could be a small back-reaction effect on the detector when a boundary is introduced, but to the order we are interested in this
is
not important.)
As for the propagation of the particle from the boundary to the detector, we already have that
taken care of in terms of $\del_{1'} G (x', x_1'; y, y_1 )$.

\section{ The effective action}

We now turn to the  action $S_\K$ for the bilocal boundary
field $\K (x,y)$. As explained earlier, our strategy is to integrate out the $\vf$-field and
identify the kind of terms which would be generated.
We then expect that such terms should exist in the action for
$\K$. The full investigation taking account of the 
bilocal nature of $\K$ would be rather involved. Here we make a first attempt by considering 
$\K$'s of the form $\K (x, y) = \kappa (x)\, \delta (x-y) =
\kappa_0 \,\delta (x-y) + \delta \kappa (x)\, \delta (x-y)$.
With this choice of $\K(x,y)$, we can easily integrate out the fields $\vf$ in the
boundary action
and
obtain the effective action for $\kappa$ as
\beq
\Delta S_{\rm eff} = {1\over 2} \, \Tr \log ( M + \kappa_0 + \delta \kappa )
\label{17}
\eeq
with $M = \sqrt {k^2}$. Here we have reverted to the Euclidean signature;
since we are embarking on loop calculations, it is easier to do this.
We are also considering the case of the entire $d$-dimensional boundary contributing; we do not assume
Dirichlet condition on any part of the boundary. (Such regions, because we need
$\kappa \rightarrow \infty$, cannot be treated in a perturbative fashion. We will need to use
a modified $M$ with $\vf$ expanded in mode functions with support only in non-Dirichlet regions.) 
Also, to minimize clutter in equations, we will use the vector notation for momenta and
coordinates; they will designate components tangential to the boundary.

Expanding (\ref{17}) in powers of $\delta \kappa$, we get the first correction as
\beq
\Delta S_{\rm eff}^{(1)} = {1\over 2} \int d^dx {d^d k \over (2 \pi )^d} 
\left[{ e^{i \vec{k} (\vec{x}-\vec{y}) }\over \sqrt{\vec{k}^2} + \kappa_0} \, \delta \kappa (x) \right]_{y\rightarrow x}
\label{18}
\eeq
The part of $\delta \kappa$ which is independent of $x$ can be absorbed into $\kappa_0$, so that, without loss of generality, we can take $\int d^dx \, \delta \kappa (x) = 0$.
Thus, we can take $\Delta S_{\rm eff}^{(1)} = 0 $.

The term which is quadratic in $\delta \kappa$ is given by
\beqar
\Delta S_{\rm eff}^{(2)} &=& - {1\over 4} \int d^d x\, d^d y\, {d^d k \over (2\pi )^d}
{d^d p \over (2\pi )^d} \, \delta \kappa (x) \,{e^{i\vec{k } (\vec{x}-\vec{y}) } e^{i \vec{p} (\vec{y}- \vec{x})} \over ( \sqrt{\vec{k}^2} + \kappa_0) (\sqrt{\vec{p}^2} + \kappa_0 )} \, \delta \kappa (y)
\nonumber\\
&=& {1\over 2} \int d^d x\, d^d y\,
\delta \kappa (x) \left[ {d^d l \over (2\pi )^d} \, e^{-i \vec{l} (\vec{x}-\vec{y})} \, f_d(\vec{l})\right] \, \delta \kappa (y)
\label{19}
\eeqar
where
\beq
f_d (\vec{l}) = - {1\over 2} \int {d^d p \over (2 \pi )^d } {1\over \left(\sqrt{(\vec{p}+\vec{l})^2}\,\, + \kappa_0 \right)\,
\left( \sqrt{\vec{p}^2} + \kappa_0\right)}
\label{20}
\eeq
The function $f_d (\vec{l})$ is even in $\vec{l}$ and we will see it has an expansion in terms of $l^2$, where $l=|\vec{l}|$. Below, we evaluate (\ref{20}) in the case $d=3,2,1$. 
Carrying out the angular integrations in (\ref{20}), we get
\beq
f_{d=3} (l) = - {1\over 8 \pi^2 l} \Biggl[ 
\int_0^{\infty} {dp \, p \over (p + \kappa_0)} \left[l+p-|l-p| - \kappa_0 \log \left( {\kappa_0 + l +p \over \kappa_0 +|l - p|} \right)\right]
\label{21}
\eeq
The above integral is divergent. This is due to the fact that $f_d(l=0)$ in (\ref{20}) diverges for $d\ge 2$. After introducing an upper cutoff $\Lambda$ for the $p$ integration we find
\beqar
f_{d=3} (l) &=& - {1\over 8 \pi^2 l} \Biggl[ 
\int_0^l {dp \, p \over (p + \kappa_0)} \left[
2 p - \kappa_0 \log \left( {\kappa_0 + l +p \over \kappa_0 +l - p} \right)\right]
\nonumber\\
&& \hskip .6in + \int_l^\Lambda {dp \, p \over (p + \kappa_0)}
\left[
2 l - \kappa_0 \log \left( {\kappa_0 + l +p \over \kappa_0 +p - l} \right)\right]
\Biggr]
\label{21a}
\eeqar
For the effective action, we are interested in a small $l/\kappa_0$-expansion.
The integrals can be evaluated in a straightforward manner to obtain
\beq
f_{d=3} (l) = -{1\over 8 \pi^2} \left[ 2 \Lambda + 2 \kappa_0 - 4 \kappa_0 \log \left({\Lambda + \kappa_0 \over \kappa_0 } \right) \right] ~+~ {1 \over 72 \pi^2 \kappa_0} \, l^2 ~+ \cdots
\label{22}
\eeq
where the ellipsis indicates terms of higher order in powers of $l^2$.
Using this expression for $f_{d=3}(l)$ in (\ref{19}), we see that the first term
of (\ref{22}) is like a mass renormalization, while the second term is like a wave function renormalization. Since the mass renormalization is divergent as
$\Lambda \rightarrow \infty$, we have to postulate that the starting action has a term
$\half \int \mu \, (\delta \kappa )^2$. Even though it is not forced by the divergence structure, we can also consider adding a term proportional to $\int (\nabla \delta \kappa )^2$.
Thus we consider the starting action $S_{0\K}$
\beq
S_{0\K}= {Z_0\over 2} \int \delta \kappa (-\nabla^2  + \mu ) \delta \kappa 
\label{23}
\eeq
With the addition of the contribution from (\ref{19}), we get the effective
action
\beqar
S_\K ^{d=3}&=& {Z\over 2} \int \delta \kappa (-\nabla^2  + \mu_{ren}  ) \delta \kappa \nonumber\\
\mu_{ren}  &=& \mu - {1\over 8 \pi^2} \left[ 2 \Lambda + 2 \kappa_0 - 4 \kappa_0 \log \left({\Lambda + \kappa_0 \over \kappa_0 } \right) \right] + \cdots\label{24}\\
Z&=& Z_0 + {1 \over 72 \pi^2 \kappa_0} + \cdots
\nonumber
\eeqar
In (\ref{24}), we could consider eliminating $Z$ by scaling $\delta \kappa$ to get a canonically normalized kinetic term for $\delta \kappa$, but this will redefine terms with higher powers
of $\delta \kappa$ in the expansion of (\ref{17}) as well as the $\half \int \delta \kappa\, \vf^2 $ term.
So we will leave the action (\ref{24}) as it is.

In the case of $d=2$ we find
\beq
f_{d=2} (l) = -{1\over 4 \pi} \left[  \log \left({\Lambda + \kappa_0 \over \kappa_0 } \right) -1\right] ~+~ {1 \over 96 \pi \kappa_0^2} \, l^2 ~+ \cdots
\label{23a}
\eeq
This implies that the corresponding renormalized parameters for $d=2$ would be
\beqar
\mu_{ren} &=& \mu - {1\over 4 \pi} \left[  \log \left({\Lambda + \kappa_0 \over \kappa_0 } \right) -1\right]+ \cdots\label{24a}\\
Z&=& Z_0 + {1 \over 96 \pi \kappa_0^2} + \cdots
\nonumber
\eeqar
Similarly in the case of $d=1$ we find
\beq
f_{d=1} (l) = -{1\over {2 \pi \kappa_0}} +{ 1 \over {12 \pi \kappa_0^3}} \, l^2 ~+ \cdots
\label{23b}
\eeq
There are no divergences in this case and the corresponding renormalized parameters are
\beq
\mu_{ren} = \mu - {1\over {2 \pi \kappa_0}} + \cdots, \hskip .2in
Z = Z_0 + {1 \over {12 \pi \kappa_0^3}} + \cdots
\label{24b}
\eeq

The calculations we have done are just the beginning in elucidating the nature of
the action for $\K$. Even with the form of $\K$ we have chosen, there are higher order corrections possible. Further the full bilocal nature of $\K$ will bring in further complications and new features.
Clearly continued investigations into these questions are needed, which we 
propose to take up in future. For now, some observations on the nature of
$K$ might be useful.

The topology of the unitary transformation in (\ref{1}) is an interesting question. It is a
well known result, namely Kuiper's theorem,
 that all the homotopy groups of
the set of unitary transformations on an infinite dimensional Hilbert space $\H$ vanish. Therefore, {\it a priori},
 it would seem that
 we have only trivial topology. However, there are operators $U$ which differ from the identity only
on a finite-dimensional subspace of the Hilbert space $\H$, or phrased differently, of the form
$U = 1 + \O$, where $\O$ is a compact operator. The homotopy groups  of such unitary operators can be nontrivial in general.
The question of which type of $U$'s will be relevant for a given physical situation
has to be answered on physical grounds such as requiring
finiteness of energy. In other words, it will be determined by $S_\K$.
If we consider restricted classes of $U$'s, we can certainly get nontrivial topology;
this has the potential to lead to interesting solitonic excitations of the boundary field.
For such cases, staying within the chosen class, one may not be able to use
simple boundary conditions such as the Dirichlet or Neumann ones as topology forces variation
of $\K (x, y)$ with $x, y$ in a certain way.
Clearly, the calculation of the action is crucial to discuss such questions. 
The case we have considered for the calculation of the action, namely,
$\K (x,y) = \kappa (x) \, \delta (x-y)$ does not quite fit into the case of
$U$'s of the form $U = 1 +\O$, with $\O$ compact.
So new techniques are needed for evaluating the action in the general case.

\section{Discussion}

Boundary conditions can in general be encoded as an integral kernel.
This kernel can be viewed as a bilocal dynamical field on the boundary.
Physical effects due to boundary conditions, such as radiation from the boundary
for time-dependent boundary conditions, can be calculated 
using standard field theory techniques such as perturbation theory.
This is illustrated in section 3. The nature of the action which governs
this bilocal field was considered in section 4.
By integrating out the bulk fields, we can obtain the general form of the action.
This is worked out for a limited class of boundary conditions
(which included the Robin case). The more general question of the action governing
the bilocal field is under investigation.

There are also more general situations than what we have considered here to which this method can evidently be applied. Moving boundaries, multiple boundaries, temperature dependence are some obvious examples. Situations which require nonlocal boundary conditions for which the bifocal nature of $\K (x,y)$ and the action governing it are fully operational would be another important case. These will be taken up in future work.

\bigskip
We thank Dan Kabat for a critical reading of the manuscript.
This research was supported in part by the U.S.\ National Science
Foundation grants PHY-1213380, PHY-1417562
and by PSC-CUNY awards.




\begin{thebibliography}{99}

\bibitem{reviews} For general reviews on the Casimir effect, see
K.A. Milton, {\it Recent developments in Casimir effect}, J. Phys. Conf. Ser. {\bf 161}, 012001 (2009);
K. A. Milton, {\it The Casimir Effect: Physical Manifestations of Zero-Point Energy} (World Scientific, 2001);
M. Bordag, U. Mohideen and V.M. Mostepanenko, Phys. Rept. {\bf 353}, 1 (2001);
M. Bordag, G.L. Klimchitskaya, U. Mohideen and V.M. Mostepanenko, {\it Advances in the Casimir Effect}
(International Series of Monographs on Physics, 2009).

\bibitem{vonNeumann}
J. von Neumann, Math. Ann. {\bf 102} 49 (1929).

\bibitem{asorey}
M. Asorey, A.Ibort and G. Marmo,  \IJMP~  {\bf A20}  1001 (2005);
M. Asorey, D. Garcia-Alvarez, J. M. Munoz-Castaneda,
J. Phys.~ {\bf A39} 6127 (2006);
M. Asorey, J. M. Munoz-Castaneda, J. Phys. ~{\bf A41} 304004 (2008).

\bibitem{bal} A.P. Balachandran {\it et al}, \IJMP~{\bf A09}, 3417 (1994);
M. Asorey, J.M. Munoz-Castaneda, \NP~{\bf B874}, 852 (2013);
M. Asorey, A.P. Balachandran and J.M. Perez-Pardo, arXiv:1505.03461[hep-th]

\bibitem{govindarajan} T.R. Govindarajan and R. Tibrewala, \PR~{|bf D83},
124045 (2011); T.R. Govindarajan and V.P. Nair, 
 \PR {\bf D89}, 025020 (2014); T.R. Govindarajan and R. Tibrewala, arXiv:1506.05243

\bibitem{KKN1}
D.~Kabat, D.~Karabali, and V.~P. Nair, \PR {\bf D8~1}, 125013 (2010); \PR~ {\bf D84}, 129901 (2011) (Erratum).

\bibitem{KKN2}
 D.~Kabat, D.~Karabali, V.~P.~Nair, \PR~{\bf D82}, 025014 (2010).
 
 \bibitem{KK} D. Kabat and D. Karabali, \PR~{\bf D84}, 065029 (2011).
 
 \bibitem{KN1} D. Karabali and V.P. Nair, \PR~{\bf D87}, 105021 (2013).
 
 \bibitem{experiment} J.R. Johansson, G. Johansson, C.M. Wilson and F. Nori, \PRL~{\bf 103}, 147003 (2009); \PR~{\bf A82}, 052509 (2010); J.R. Johansson, G. Johansson, C.M. Wilson, P. Delsing and F. Nori, \PR~{\bf A87}, 1234 (2013).
 
 \bibitem{SF} H.O. Silva and C. Farina, \PR~{\bf D84}, 045003 (2011).
 
  \bibitem{Rego} A. Rego, J.P.S. Alves, D.T. Alves and C.Farina, \PR~{\bf A88}, 032515 (2013).
 
 \bibitem{fosco} C.D. Fosco, F.C. Lombardo and F.D. Mazzitelli, \PR~{\bf D87}, 105008 (2013).
 

\bibitem{cavalcanti} L.C.~de~Albuquerque and R.M.~Cavalcanti, J.Phys. {\bf A37}  7039 (2004). 








\end{thebibliography}
\end{document}